\documentclass[aps,twocolumn,superscriptaddress]{revtex4-1}
\usepackage{epstopdf}
\usepackage{graphicx}
\usepackage{dcolumn}
\usepackage{bm,bbm,bbold}
\usepackage{color,xcolor, soul}
\sethlcolor{green}
\usepackage{amsfonts}
\usepackage{amsmath}
\usepackage{braket}
\usepackage[normalem]{ulem}
\usepackage{mathrsfs}   
\usepackage[percent]{overpic}
\usepackage[colorlinks=true,citecolor=blue]{hyperref}
\usepackage{hyperref}
\hypersetup{colorlinks=true,citecolor=blue,linkcolor=blue,urlcolor=blue}

\def\kbar{$\bar{\mathrm{K}}$}

\def\kbarp{$\bar{\mathrm{K}}^{\prime}$}

\def\WS2{WS$_2$}
\def\MoS2{MoS$_2$}{

\begin{document}
\title{Layer and orbital interference effects in photoemission from transition metal dichalcogenides}

\author{Habib Rostami}
\email[Electronic address: ]{habib.rostami@su.se}
\affiliation{Nordita, Center for Quantum Materials, KTH Royal Institute of Technology and Stockholm University, Roslagstullsbacken 23, SE-106 91 Stockholm, Sweden}
\author{Klara Volckaert}
\author{Nicola Lanata}
\author{Sanjoy K. Mahatha}
\affiliation{Department of Physics and Astronomy, Interdisciplinary Nanoscience Center, Aarhus University,
8000 Aarhus C, Denmark}
\author{Charlotte E. Sanders}
\affiliation{Central Laser Facility, STFC Rutherford Appleton Laboratory, Harwell 0X11 0QX, United Kingdom}
\author{Marco Bianchi}
\affiliation{Department of Physics and Astronomy, Interdisciplinary Nanoscience Center, Aarhus University,
8000 Aarhus C, Denmark}
\author{Daniel Lizzit}
\author{Luca Bignardi}
\thanks{current address: Department of Physics, University of Trieste, Via Valerio 2,Trieste 34127, Italy}
\author{Silvano Lizzit}
\affiliation{Elettra-Sincrotrone Trieste, S.S. 14 Km 163.5, Trieste 34149, Italy}
\author{Jill A. Miwa}
\affiliation{Department of Physics and Astronomy, Interdisciplinary Nanoscience Center, Aarhus University,
8000 Aarhus C, Denmark}
\author{Alexander V. Balatsky}
\affiliation{Nordita, Center for Quantum Materials, KTH Royal Institute of Technology and Stockholm University, Roslagstullsbacken 23, SE-106 91 Stockholm, Sweden}
\author{Philip~Hofmann}
\affiliation{Department of Physics and Astronomy, Interdisciplinary Nanoscience Center, Aarhus University,
8000 Aarhus C, Denmark}
\author{S{\o}ren~Ulstrup}
\email[Electronic address: ]{ulstrup@phys.au.dk}  
\affiliation{Department of Physics and Astronomy, Interdisciplinary Nanoscience Center, Aarhus University,
8000 Aarhus C, Denmark}
\begin{abstract}
In this work, we provide an effective model to evaluate the one-electron dipole matrix elements governing optical excitations and the photoemission process of single-layer (SL) and bilayer (BL) transition metal dichalcogenides. By utilizing a $\bm k\cdot\bm p$ Hamiltonian, we calculate the photoemission intensity as observed in angle-resolved photoemission from the valence bands around the \kbar-valley of MoS$_2$. In SL MoS$_2$ we find a significant masking of intensity outside the first Brillouin zone, which originates from an in-plane interference effect between photoelectrons emitted from the Mo $d$ orbitals. In BL MoS$_2$ an additional inter-layer interference effect leads to a distinctive modulation of intensity with photon energy. Finally, we use the semiconductor Bloch equations to model the optical excitation in a time- and angle-resolved pump-probe photoemission experiment. We find that the momentum dependence of an optically excited population in the conduction band leads to an observable dichroism in both SL and BL MoS$_2$.
\end{abstract}
\date{\today}
\maketitle
\section{Introduction}

Single-layer (SL) semiconducting transition metal dichalcogenides (TMDCs) belong to the D$_{\rm 3h}$ symmetry group, which implies the presence of a three-fold rotation symmetry and the absence of spatial inversion symmetry. Semiconducting SL TMDCs can be realized by the composition MX$_2$, where the transition metal M=\{Mo,W\} is sandwiched between layers of the chalcogen X=\{S,Se\} in a trigonal prismatic structure. The combination of a non-centrosymmetric lattice and heavy transition metals leads to a large spin-splitting in the valence band (VB) \cite{Xiao:2012ab}, valley-selective optical excitations \cite{zengvalley2012,makcontrol2012,Cao_2012}, a valley-Zeeman effect \cite{Srivastava_2015,Aivazian:2015} and a pronounced second-harmonic generation effect \cite{Kumar_2013}. When stacking two trigonal prismatic SLs to form a bilayer (BL) TMDC, electrostatic repulsion between the anions leads to a structure in which the two layers are rotated by 180$^{\circ}$ against each other, giving rise to an inversion centre between the layers. In this so-called 2H structure, all bands are spin-degenerate and the spin- and valley-degrees of freedom are no longer accessible, unless the inversion symmetry is broken by  a supporting substrate, electrical gating or selective probing of individual layers \cite{Yuan:2013,Gong:2013,Jones:2014a,Riley:2014,Riley:2015aa,Razzoli:2017}.   

The key electronic properties of SL and BL TMDCs are completely specified by the low energy electronic states at the \kbar~(\kbarp) corner of the Brillouin zone (BZ) described by the Bloch waves $|\psi_j({\bm q},\tau_z,s_z) \rangle$, where $j$ is a band index and $\tau_z = \pm 1$ and $s_z = \pm 1$ are the associated valley and spin indices. The wavevector ${\bm q} = {\bm k} - \tau_z {\bm K}$ describes the states around the valley points ($q\ll K$) where the VB and conduction band (CB) states are separated by a direct band gap \cite{Xiao:2012ab}. The CB derives from the Mo (W)  $d_{z^2}$ orbital, while the VB is mainly composed of $d_{x^2 -y^2}$ and $d_{xy}$ orbitals \cite{kormanyos2015} and is characterized by a significant trigonal warping effect that plays an important role in the optical, electrical and magnetic properties of the materials \cite{Kormanyos:2013,Rostami:2013,Gong:2013}.

\begin{figure}
\begin{center}
\includegraphics[width=0.49\textwidth]{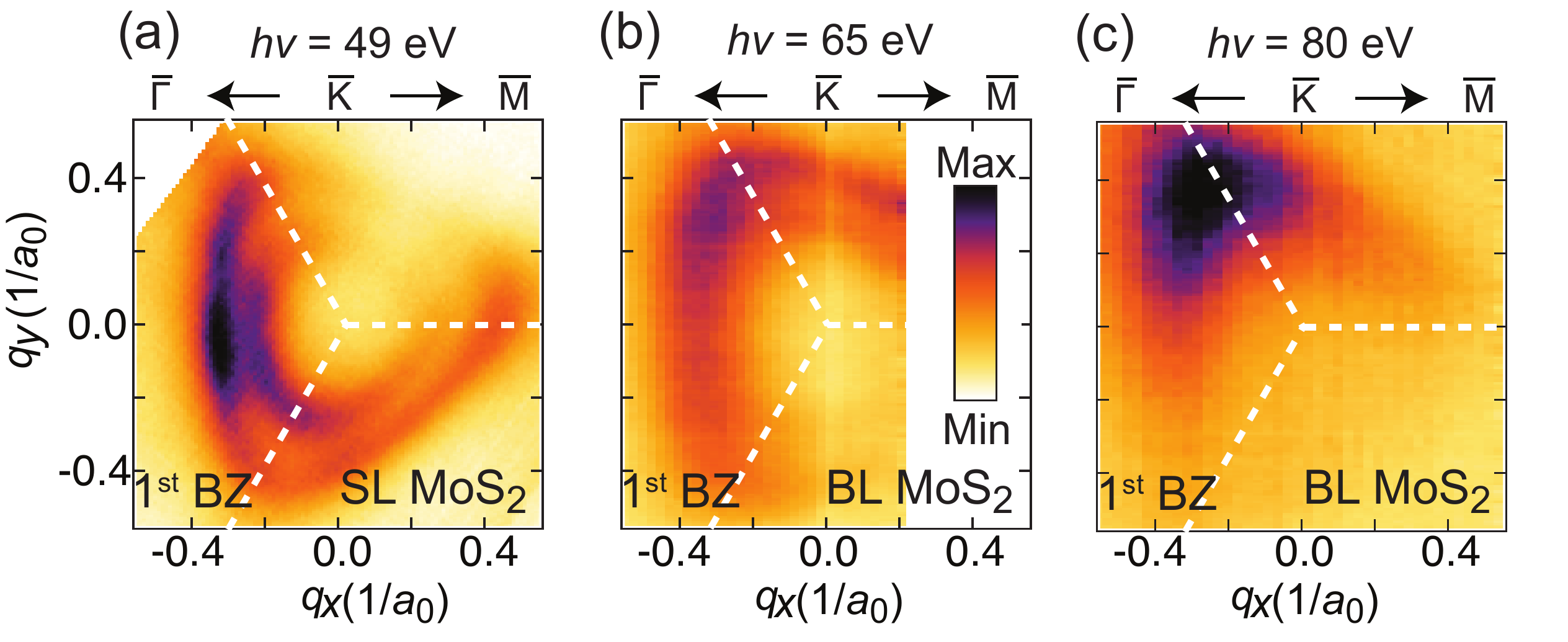}
\caption{(a)-(c) ARPES intensity at a fixed binding energy 0.24 eV below the VBM at~\kbar~for (a) SL MoS$_2$ on Au(111)  \cite{Miwa:2015aa} and (b)-(c) BL MoS$_2$ on Ag(111) \cite{ourPRL}. The dashed white lines indicate the BZ boundaries. The data were obtained using $p$-polarized photons with the given photon energies at the SGM3 beamline of the ASTRID2 synchrotron light source in Aarhus, Denmark \cite{Hoffmann:2004aa}.}
\label{fig:expFig}
\end{center}
\end{figure}

In this Article, we model the ${\bm q}$-dependent photoemission and optical selection rules emerging from the orbital, spin, valley and layer degrees of freedom from the VB and CB states in SL and BL TMDCs. This study is motivated by recent angle-resolved photoemission spectroscopy (ARPES) experiments on SL and BL MoS$_2$, as presented in Fig. \ref{fig:expFig}. Further details about the experiments are provided in Refs. \cite{Miwa:2015aa,ourPRL}. The measured photocurrent in ARPES is given by ${\cal I}_n(E,{\bm q}) =  |{\cal M}_n(E,{\bm q})|^2 {\cal A}_n(E,{\bm q}) f_{FD}(E)$. Here, ${\cal A}_n$ stands for the photohole spectral function, $f_{FD}$ is the Fermi-Dirac function and ${\cal M}_n \propto \langle {\psi}_f | \hat{{\bm \epsilon}}\cdot\hat{{\bm p}} |\psi_i \rangle $ is the one-electron dipole matrix element \cite{Hufner:2003}. ${\cal M}_n$ describes the coupling of the initial state $|\psi_i \rangle$ in the TMDC to a free electron final state $|{\psi}_f \rangle$ via the momentum operator $\hat{{\bm p}}$ and an incident electric field with polarization $\hat{{\bm \epsilon}}$. The SL MoS$_2$ photocurrent measured with 49 eV photons is presented as a constant energy cross section, 0.24~eV below the VB maximum (VBM) at \kbar~in Fig.~\ref{fig:expFig}(a). Two trigonally warped contours corresponding to the spin-orbit split VBs are visible. A strong variation of the photoemission intensity, with the highest intensity observed in the first BZ, indicates a pronounced momentum dependence of the photoemission matrix elements. Constant energy contours extracted at the same energy below the VBM at \kbar~for BL MoS$_2$ exhibit similar trigonal features in addition to a redistribution of intensity around the edge of the first BZ depending on the photon energy as seen in Figs.~\ref{fig:expFig}(b)-(c). 

The momentum-dependence of the matrix elements encodes the symmetry and orbital character of the initial state. Moreover, despite the extended nature of the initial state,  interference between photoelectrons emitted from different sites in the unit cell can play an important role. For example, in graphene the two carbon basis atoms of the primitive unit cell cause a sublattice interference effect that modulates the photoemission intensity of the $\pi$-states composed from the $p_z$ orbitals centered on the carbon atoms \cite{Shirley:1995aa,mucha_prb_2008,Lizzit:2010aa}. The intensity modulation due to the matrix elements can be strongly dependent on the energy and polarization of the photon beam \cite{Gierz:2011,Liu:2011,Gierz:2012}. This has been  exploited for more complex layered systems such as the prototypical topological insulator Bi$_2$Se$_3$ to extract information about orbital angular momentum and spin texture \cite{Wang:2011s,Jung:2011,Park:2012}. However, the roles of photoemission geometry, layer-dependent dispersion and the electron final state are highly non-trivial to disentangle from these initial state effects \cite{Ishida:2011,Zhu:2013aa,Scholz:2013,Zhu:2014ac,Xu:2015}. A careful evaluation of the matrix element effects is therefore a crucial part of such analysis.

Here, we show that the intensity modulations seen in Fig. \ref{fig:expFig} for SL and BL MoS$_2$ originate from intra- and inter-layer interference effects between photoemitted electrons from the Mo $d$ orbitals. We apply a semi-analytical approach based on low-energy massive Dirac-like models around the direct band gap in both SL and BL MoS$_2$ in order to calculate the ARPES intensity of the VB and CB states. Since the CB is usually unoccupied it is necessary to utilize an optical excitation to populate these states and then collect the photoemission intensity from the resulting excited state population, which is possible in an ultrafast time-resolved (TR) ARPES experiment \cite{Antonija-Grubisic-Cabo:2015aa,Hein:2016,Bertoni:2016,Ulstrup2017,Beyer:2019aa}. We therefore employ the semiconductor Bloch equations to calculate the excited state population in the CB minimum (CBM), and show that the associated intensity is strongly dependent on ${\bm q}$ and pump pulse polarization. 

The purpose of this paper is to present a tractable model that can be used as a framework to evaluate the role of matrix element effects in ARPES and TR-ARPES experiments performed on TMDCs. The rest of this paper is organized into three sections. In Section II we provide an analytical analysis of the dipole matrix elements associated with photoemission from the SL and BL MoS$_2$ VBs and CBs. In section III we extend this analysis to include optical excitation and free carrier population of CB states. Our results are summarized in Section IV. 

\section{Matrix elements for the photoemission process}

In this section we adapt the procedure in Ref. \citenum{moser_2017} in order to determine ${\cal M}_n$ for the TMDCs. The initial state is written as a Bloch wave $|\psi_i\rangle=|\psi_j({\bm q},\tau_z,s_z) \rangle$ and the final state is approximated as a single plane wave $|\psi_f\rangle =|\bm k_f\rangle $, leading to 
\begin{align}
{\cal M}_n({\bm q},\tau_z,s_z)&=\hat{{\bm \epsilon}}\cdot{\bm k}_f \langle {\bm k}_f |\psi_n({\bm q},\tau_z,s_z) \rangle, 
\end{align}
where $\bm k_f$ is the wavevector of the photoemitted electron. In an ARPES experiment the direction $\hat {\bm k}_f = {\bm k_f}/k_f = (\theta_f,\phi_f)$ is given by the measurement of polar and azimuthal emission angles $\theta_f$ and $\phi_f$. The magnitude $k_f$ is obtained from the kinetic energy $E_k$ of the photoelectron. Due to translational symmetry, the in-plane momentum is conserved in the photemission process (modulo a reciprocal lattice vector)  ${\bm k}_{f\parallel} = \tau_z {\bm K} + {\bm q}$. The situation is less simple for the out-of-plane momentum. This is strictly not a good quantum number for an electron near a surface, especially given the short inelastic mean free path of the photoelectrons. However, when we simply consider a free electron final state inside the solid, the coupling to the photoemitted electron outside the surface only requires that an energy scale shift in the form of the inner potential $V_0$ be taken into account, resulting in a refraction at the surface barrier. We apply this procedure here and write the perpendicular momentum \emph{of the initial state} as $k_{f\perp} = \left(2m_0(E_{k}\cos^2\theta_f + V_0)\right)^{1/2}/\hbar$. Here, $m_0$ is the free electron mass and $\hbar$ is Planck's constant.  $E_k = h\nu - E_{bin} - \Phi$ is the electron kinetic energy at the given photon energy $h\nu$, work function $\Phi$ and binding energy $E_{bin}$ of the initial state. A variation of $h\nu$ therefore implies a change of  $k_{f\perp}$. Note that this treatment is designed to handle the photoemission process for semi-infinite solids. In a SL or BL, the meaning of $V_0$ and its consequences are less clear \cite{Winkler:2017aa} and the particular choice of $V_0$ does not have a significant qualitative effect on the results of this paper. The reader may choose to interpret $k_{f\perp}$ as a measure of the photon energy (for states of a given binding energy) and remember that a comparison of calculated and measured cross-section variations as a function of $h\nu$ may require the adjustment of $V_0$. 

\begin{figure}
\begin{center}
\includegraphics[width=0.49\textwidth]{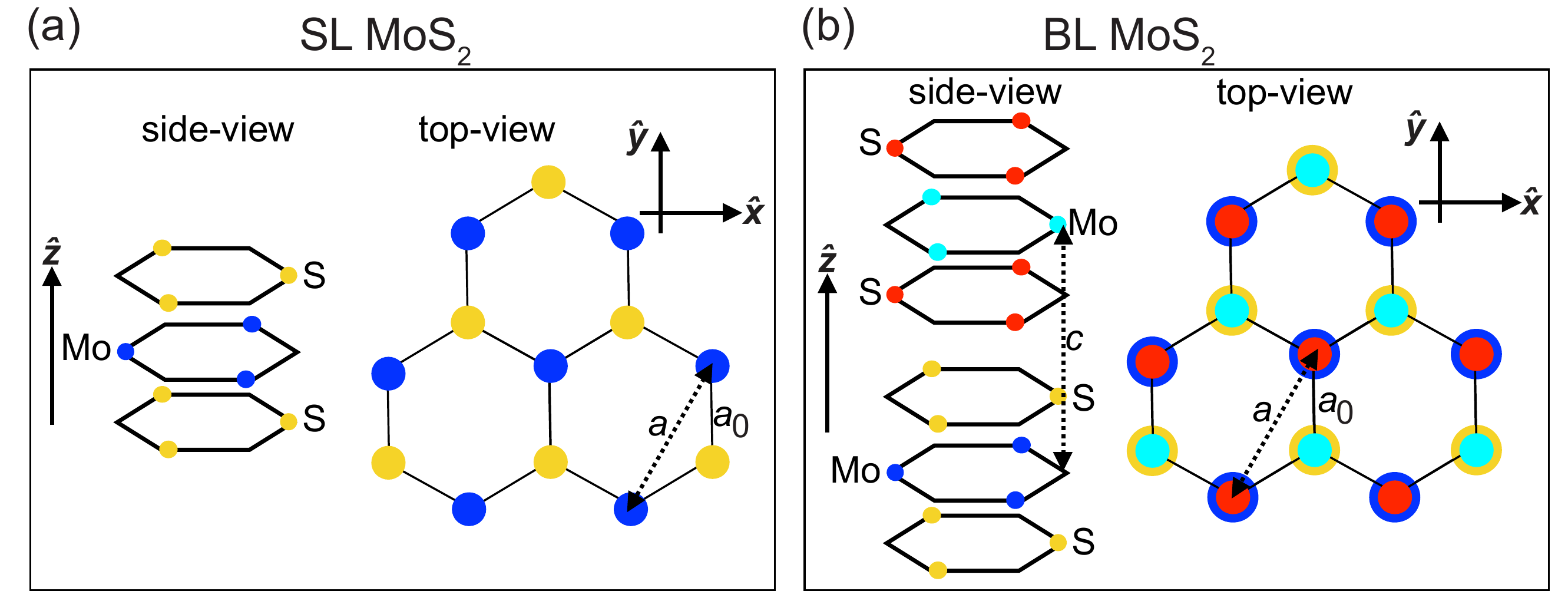}
\caption{(a)-(b) Schematic of (a) SL and (b) BL MoS$_2$ structures. The lattice constant  $a\sim 3.16\rm$~\AA~in the $\{\hat{\bm{x}},\hat{\bm{y}}\}$-plane is indicated by a double-headed arrow in the top-view. The inter-layer separation $c = 7.0$~\AA~along $\hat{\bm{z}}$ is indicated by a double-headed arrow in the side-view in (b). The side-length of a hexagonal unit is indicated as $a_0 = a/\sqrt{3}$.}
\label{fig:bilayer_mx2}
\end{center}
\end{figure}

In the following we consider the geometry of SL and BL TMDCs presented in Figs. \ref{fig:bilayer_mx2}(a)-(b) where the ``zig-zag" direction of the lattice is oriented along $\hat {\bm x}$ and the ``arm-chair" direction is oriented along $\hat {\bm y}$. This orientation leads to valley points at the wavevector $ \bm k=\tau_z \bm K$ where  $ \bm K = \hat {\bm x} 4\pi/3a $, with $a\sim 3.16\rm$~\AA~being the lattice constant in MoS$_2$. In the case of BL MoS$_2$, we emphasize that there are two Mo basis atoms located at $\bm u$ and $\text{-}\bm u$ where $2\bm u = a_0 \hat {\bm y} + c \hat {\bm z}$ with $a_0 = a/\sqrt{3}$ and $c = 7.0$~\AA~is the separation between neighboring Mo planes \cite{Wu:2013}.

\subsection{SL TMDC}

The initial Bloch state at a given ${\bm k}$ and spin index is written as a linear combination of atomic orbitals
\begin{align}
&|\psi_{j}({\bm k},s_z) \rangle =
\sum_{\bm R}\sum_{n\ell m}  \frac{e^{i{\bm k} \cdot {\bm R}} }{\sqrt{N}} U^{j}_{n\ell m}({\bm k},s_z)  |{\bm R};n\ell m\rangle,
\end{align}
where $N$ stands for the number of lattice sites, the sum $\sum_{\bm R} $ is evaluated over all lattice sites, $\{n,\ell,m\}$ are the atomic orbital quantum numbers and the ${\bm k}$-dependent lattice periodic function  $U^j_{n\ell m}$ is specified in further detail below. We utilize the relation
\begin{align}
&\langle {\bm k}_f|{\bm R}; n\ell m \rangle
=\langle {\bm k}_f|e^{-i{\bm k}\cdot{\bm R}}|{\bm 0}; n\ell m \rangle
 \nonumber\\&
 = e^{-i{\bm k}_f\cdot {\bm R}} \langle {\bm k}_f|{\bm 0};n\ell m \rangle = \Phi_{n\ell m}({\bm k}_f) e^{-i{\bm k}_f\cdot {\bm R}},
\end{align}
where the Fourier transform of the atomic orbital is given by $\Phi_{n\ell m} ({\bm k}_f)   =  f_{n\ell}(k_f) Y_{\ell m}(\theta_f,\phi_f)$ with  \cite{podolsky_pr_1929,moser_2017}
\begin{align}
  f_{n\ell}(x)= {\cal N}_{n\ell} \frac{(-i x)^\ell}{(x^2+1)^{\ell+2}} C^{(\ell+1)}_{n-\ell-1}\left(\frac{x^2-1}{x^2+1} \right),
\end{align}
where $C^{(\alpha)}_{n}(x)$ are Gegenbauer polynomials, ${\cal N}_{n\ell}$ is a numerical pre-factor~\cite{podolsky_pr_1929}, and $Y_{\ell m}$ are spherical harmonics. 
The in-plane momentum conservation is satisfied by $\sum_{\bm R} e^{i{\bm R}\cdot({\bm k}-{\bm k}_f)}= N\delta_{{\bm k} , {\bm k}_{f\parallel}}$, which implies  
\begin{align}
\langle {\bm k}_f |\psi_{j}({\bm k},s_z) \rangle = 
&\sqrt{N}\delta_{{\bm k} , {\bm k}_{f\parallel}} \notag \\
&\times \sum_{n\ell m} f_{n\ell}(k_f) U^{j}_{n\ell m}({\bm k},s_z)  Y_{\ell,m}(\hat{\bm k}_f).
\end{align}
We now state this expression for the valley points, which are composed of Mo $d$ orbitals in the fourth ($n=4$) principle quantum shell. Accordingly, we only consider  $|A\rangle\equiv d_{z^2}\equiv |2,0\rangle $ and $|B\rangle\equiv(d_{x^2-y^2}- i \tau_z d_{xy})/\sqrt{2}\equiv |2,-2\tau_z\rangle$~\cite{Xiao:2012ab}. Therefore, for a given valley index, $\tau_z$, we find  
\begin{align}
&\langle {\bm k}_f |\psi_{j}({\bm q},\tau_z,s_z) \rangle = 
\sqrt{N}
\delta_{{\bm k} , {\bm k}_{f\parallel}}  f_{42}(k_f) 
\Big\{
U^{j}_{42,0}({\bm q},\tau_z,s_z)\times  
\nonumber\\&Y_{2,0}(\hat{\bm k}_f) 
+
U^{j}_{42,-2\tau_z}({\bm q},\tau_z,s_z)  Y_{2,-2\tau_z}(\hat{\bm k}_f) 
\Big\}.
\end{align}
Note that $ U^{j}_{n\ell m}({\bm q},\tau_z,s_z) \equiv U^{j}_{n\ell m}(\tau_z {\bm K}+{\bm q},s_z)$.
In order to estimate the value of  $U^{j}_{n\ell m}$ we consider a ${\bm k} \cdot {\bm p}$ Hamiltonian of a SL TMDC in the basis of $\{|A\rangle,|B\rangle\}$ orbitals, which is given by
\begin{equation}\label{eq:Ham_SL}
\hat{\cal H}_{\rm SL}({\bm q},\tau_z,s_z)   =   \hat{\cal H}_{\rm iso}({\bm q},\tau_z)+\hat{\cal H}_{\rm tw}({\bm q},\tau_z)
 +  \hat{\cal H}_{\rm soc} \tau_z s_z.
\end{equation}
The spin-orbit coupling term reads $ \hat{\cal H}_{\rm soc}  =  \lambda_I \hat I+\lambda \hat \sigma_z $, where $\hat I$ is the identity matrix, $\hat{\sigma}_{\alpha}$ is the Pauli spin matrix in the direction $\alpha$, and the spin-orbit coupling parameters are given in terms of $\lambda_c=\lambda_I+\lambda\approx-5.5$~meV and $\lambda_v=\lambda_I-\lambda\approx 74.5$~meV, which are the spin-orbit coupling in the CB and VB, respectively. The isotropic, $\hat {\cal H}_{\rm iso}$, and trigonal warping, $\hat {\cal H}_{\rm tw}$, terms are given by the following two-band Hamiltonians \cite{Rostami:2013,Kormanyos:2013,Rostami:2016}  
\begin{align}
\hat {\cal H}_{\rm iso}({\bm q},\tau_z) &= \Delta\hat\sigma_z +
 a_0 t_1(\tau_z q_x \hat\sigma_x +q_y\hat\sigma_y) 
\nonumber\\&
+a^2_0(\alpha\hat I+ \beta\hat \sigma_z )q^2,
\nonumber\\
\hat{\cal H}_{\rm tw}({\bm q},\tau_z) &= a^2_0 t_2 \{(q^2_x-q^2_y) \hat\sigma_x  + 2\tau_z q_x q_y  \hat\sigma_y\}    
\nonumber\\&
+a^3_0\tau_z  (\alpha'\hat I+\beta'\hat \sigma_z) (q^3_x-3q_x q^2_y).
\end{align}

Considering electron and hole effective masses, we obtain $\alpha= \hbar^2/4\mu' a^2_0$ and $\beta = E_0 - t^2_1/E_g$ with $E_0=\hbar^2/4\mu a^2_0$ where $\mu=m_e m_h/(m_h+m_e)\approx 0.2 m_0$ and $\mu'=m_e m_h/(m_h-m_e)\approx 2.3 m_0$. Here we have used $m_e\approx0.37m_0$ and $m_h\approx0.44m_0$ for the electron and hole effective mass in SL MoS$_2$, respectively~\cite{peelaers_prb_2012,cheiwchanchamnangij_prb_2012}. Notice that $2\Delta= E_g-\lambda_c+\lambda_v$ in which $E_g=1.95$~eV \cite{Antonija-Grubisic-Cabo:2015aa} is the energy gap. For the intralayer effective hopping we set $t_1 = 2.0$~eV. The trigonal warping parameters are set to $t_2\approx-0.14$~eV, $\alpha'\approx 0.44$~eV and $\beta'\approx -0.53$~eV \cite{rostami_prb_2015,Rostami:2016}.  
Using the parameterization as $\hat{\cal H}_{\rm SL} = h_I\hat I+h_x\hat \sigma_x+h_y\hat \sigma_y+h_z\hat \sigma_z$ we obtain 
\begin{align}\label{eq:A_B}
&A_{n} \equiv U^{n}_{42,0}=
\frac{h_{x}- i  h_{y}}{\sqrt{h_{x}^2+ h_{y}^2 +  ( n | {\bm h}|-h_{z} )^2}}~,
 \nonumber\\
&B_{n} \equiv U^{n}_{42,-2\tau_z}=
\frac{n | {\bm h}|-h_{z}}{\sqrt{h_{x}^2+ h_{y}^2 +  (n | {\bm h}|-h_{z})^2}}~. 
\end{align}
Notice that for the shorthand notation we have dropped the argument $(\bm q,\tau_z,s_z)$ in the above relations. We then arrive at the following expression for the matrix element around the valley points in a SL TMDC: 
 \begin{align}\label{eq:SLTMn}
&{\cal M}_n(k_{f\perp};{\bm q},\tau_z,s_z)=
\sqrt{N}
\hat {\bm \epsilon}\cdot  {\bm k}_f  f_{42}(k_f)\times
\\&
\Big\{A_n({\bm q},\tau_z,s_z)  Y_{2,0}(\hat{\bm k}_f) 
+ 
B_n({\bm q},\tau_z,s_z) Y_{2,-2\tau_z}(\hat{\bm k}_f)  \Big \}.
\nonumber
\end{align}

\subsection{BL TMDC}
In the BL TMDC model we utilize a four-band ${\bm k}~\cdot~{\bm p}$ Hamiltonian expressed in the basis \{$c_+$, $v_+$, $c_-$, $v_-$\} where $c_+(v_+)$ labels CB (VB) of top layer and $c_-(v_-)$ labels CB (VB) of bottom layer, leading to \cite{Gong:2013,Kormanyos:2018}

\begin{equation}\label{eq:BL_4b}
\hat{\cal H}_{\rm BL}({\bm q},\tau_z,s_z)  = \begin{bmatrix} \hat{\cal H}_{-} & \hat{\cal H}_\perp \\[5pt] \hat{\cal H}^\dagger_\perp & \hat{\cal H}_{+} \end{bmatrix}
\end{equation}
with $\hat{\cal H}_{\pm} = \hat{\cal H}_{\rm SL}(\pm{\bm q},\pm\tau_z,s_z)$ and  the interlayer coupling reads 
\begin{equation}
\hat{\cal H}_{\perp}  = \begin{bmatrix} t'_{\perp} (\tau_z q_x -i q_y) &0 \\[10pt] 0 & t_{\perp}\end{bmatrix},
\end{equation}
in which $t_\perp=0.045$~eV and $t'_\perp=0.0387$~eV quantify the strength of interlayer tunneling of electrons and holes, respectively~\cite{Gong:2013,Kormanyos:2018}.  
Moreover, for the BL case we set $E_g=1.9$~eV  and $2\Delta=E_g-\lambda_c+\sqrt{t^2_\perp+\lambda^2_v}$.  
\par
The Bloch function is written as a linear combination of atomic orbitals localized on the Mo lattice sites, neglecting as usual the S atoms because their contribution to the relevant states at $\tau_z {\bm K}/-\tau_z {\bm K}$ on the top/bottom layer is negligible:
\begin{align}
|\psi_j({\bm k},\tau_z,s_z) \rangle &= \frac{1}{\sqrt{N}} \sum_{\bm R} e^{i{\bm k}\cdot {\bm R}}
\bigg\{e^{i {\bm k} \cdot {\bm u}} |{\bm R}+{\bm u,\tau_z,s_z,j}\rangle 
\nonumber\\&+ 
e^{-i {\bm k} \cdot {\bm u}}  |{\bm R}-{\bm u},\tau_z,s_z,j\rangle
\bigg\}.
\end{align}

The localized orbital in each layer can be written in terms of atomic orbitals of the transition metal (\textit{i.e.} $|n\ell,m\rangle \in \{|42,0\rangle,|42,2\rangle,|42,-2\rangle\} $):
\begin{align}
 |{\bm R}\pm{\bm u},\tau_z,s_z,j\rangle &= 
 A^{\pm}_j({\bm k},s_z)  |{\bm R}\pm{\bm u};42,0\rangle 
 \nonumber\\&+ 
 B^{\pm}_j({\bm k},s_z) \left |{\bm R}\pm{\bm u}; 42,\pm2\tau_z\right \rangle,
\end{align}
where $+/-$ corresponds to the top/bottom layer.
After taking in-plane momentum conservation into account, we obtain the following result for the matrix element in a BL TMDC: 
 \begin{align}\label{eq}
{\cal M}_n(k_{f\perp};{\bm q},\tau_z,s_z) &=
e^{-i \frac{c k_{f\perp}}{2}} {\cal M}^{+}_n(k_{f\perp};{\bm q},\tau_z,s_z) 
\nonumber\\&
+ e^{i \frac{c k_{f\perp}}{2}} {\cal M}^{-}_n(k_{f\perp};{\bm q},\tau_z,s_z)
\end{align}
in which we have 
\begin{align}
&{\cal M}^{\pm}_n(k_{f\perp};{\bm q},\tau_z,s_z)  = \sqrt{N}
\hat {\bm \epsilon}\cdot  {\bm k}_f  f_{42}(k_f) \times
\\&
\Big\{
A^{\pm}_n({\bm q},\tau_z,s_z) Y_{2,0}(\hat {\bm k}_f)
 + 
B^{\pm}_n({\bm q},\tau_z,s_z) Y_{2,\pm2\tau_z}(\hat {\bm k}_f)
\Big\}.
\nonumber
\end{align}
Hence, we use the four-band model given in Eq.~(\ref{eq:BL_4b}) and evaluate the $A^{\pm}_n$ and $B^{\pm}_n$ factors. Owing to the normalization of the eigenvectors, we have $\sum_{\ell=\pm} \{|A^{\ell}_n|^2+|B^{\ell}_n|^2\}=1$.

\subsection{Momentum-dependence of the matrix elements}
\begin{figure}
\begin{center}
\includegraphics[width=0.49\textwidth]{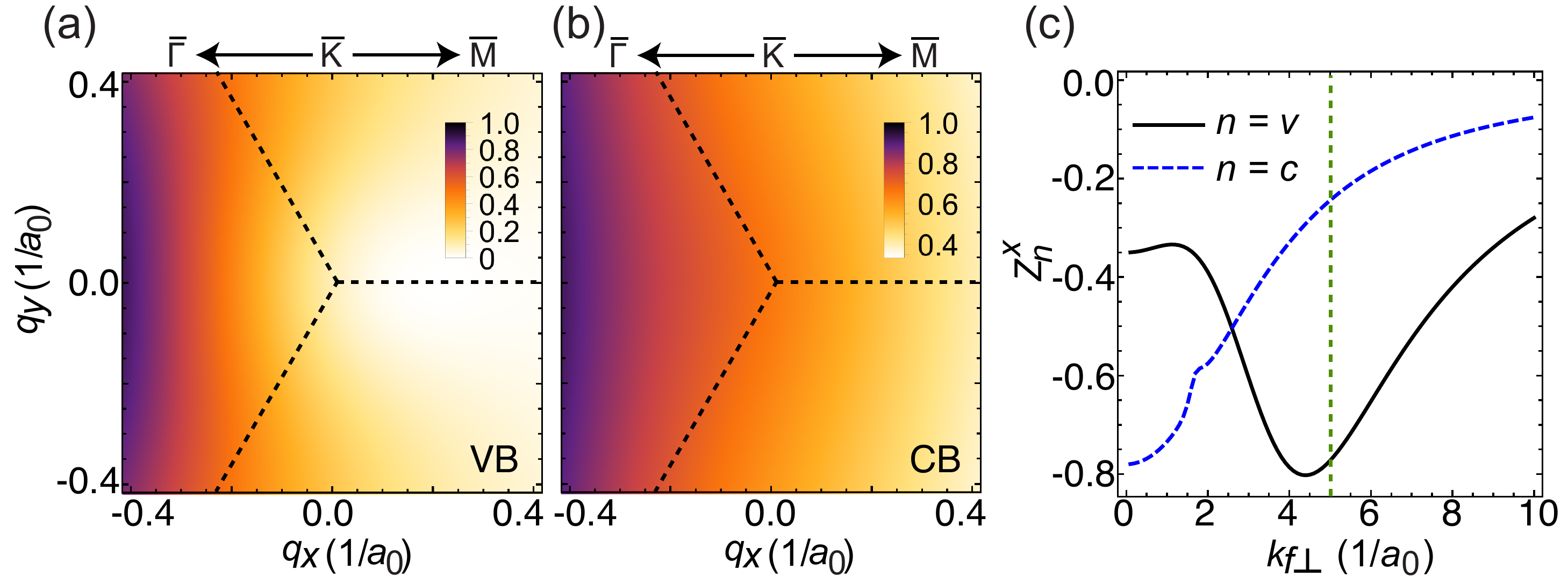}
\caption{(a)-(b) Color map of $|{\cal M}_n|^2$ in SL MoS$_2$ in the (a) VB and (b) CB for $k_{f\perp}=5/a_0$. The magnitude of $|{\cal M}_n|^2$ has been normalized to the maximum value in the plotted range. (c) Masking parameter $Z_n^x$ as a function of $k_{f\perp}$ for both VB ($n=v$) and CB ($n=c$). We set $\tau_z=s_z=1$ and $q_c=0.5/a_0$. The value of $k_{f\perp}$ used for the plots in (a)-(b) has been marked by a vertical dashed line.}
\label{fig:probe_sl}
\end{center}
\end{figure}
The numerical results for  $|{\cal M}_n|^2$ in the VB and CB of SL MoS$_2$ around \kbar~are shown in Figs.~\ref{fig:probe_sl}(a) and \ref{fig:probe_sl}(b), respectively. A monotonic decrease in magnitude is seen from negative to positive $q_x$-values while the variation in the $q_y$-direction is minor in this geometry. We will refer to the decline in $|{\cal M}_n|^2$ towards the outer BZs as a masking of the intensity.
In order to quantify this masking effect in an average manner, we introduce the following {\it masking parameter}:  
\begin{align}\label{eq:masking}
Z^\alpha_n(k_{f\perp},\tau_z,s_z)
= \frac{\sum_{\bf q} {\rm sign}(q_\alpha) |{\cal M}_n(k_{f\perp};{\bm q},\tau_z,s_z)|^2}{\sum_{\bf q} |{\cal M}_n(k_{f\perp};{\bm q},\tau_z,s_z)|^2}.
\end{align} 
The above summations are carried out in a square region with side lengths $2q_c=1.0/a_0$ centered at \kbar. Here, $\alpha$ labels the coordinate and $n$ labels the band ($v$ for VB and $c$ for CB). If $Z^x_n=Z^y_n=0$ there is no masking and thus a uniform intensity along both $q_x$ and $q_y$. A situation where $Z^x_n<0$ corresponds to a decrease of $|{\cal M}_n|^2$ for $q_x>0$, which leads to masking of the intensity towards the outer BZ. In Eq. (\ref{eq:SLTMn}) it is implicit that the matrix element depends on $k_{f\perp}$ and will thus vary with the photon energy used in an ARPES experiment. We quantify this variation by evaluating $Z^x_n$ at different $k_{f\perp}$ as shown in Fig.~\ref{fig:probe_sl}(c). In all cases we find that $Z^x_n<0$ and that $Z^y_n=0$ for the VB and CB of SL MoS$_2$ corresponding to a decrease in ARPES intensity towards the outer BZ, as explained above. The masking in the CB gets weaker with increasing $k_{f\perp}$, and in the VB it exhibits a more complex behavior with a minimum at $k_{f\perp} \approx 4.3/a_0$.

The square modulo of the SL MoS$_2$ matrix element given in Eq. (\ref{eq:SLTMn}) can be decomposed as $|{\cal M}_n|^2=P_A+P_B+P_{AB}$ where $P_A\propto |A_n|^2 |Y_{2,0}(\hat {\bm k}_f)|^2$, $P_B\propto |B_n|^2 |Y_{2,-2}(\hat {\bm k}_f)|^2$ and  $P_{AB}\propto A_n^\ast B_n Y_{2,0}(\hat {\bm k}_f)Y_{2,-2}(\hat {\bm k}_f)+c.c$. Note that the $A_n$ and $B_n$ parameters are given in Eq.~(\ref{eq:A_B}). All three terms contribute to the masking effect. This is substantially different from the case of graphene where it mainly originates from the interference term, i.e. $P^{\rm graphene}_{AB}\propto \mp (q_x/q)$ \cite{moser_2017,mucha_prb_2008} for the CB ($+$) and VB ($-$). For graphene this term originates from the two $p_z$ orbitals localized on the two carbon atoms in the primitive unit cell. In the case of SL MoS$_2$ the effect emerges from the $|A\rangle\equiv d_{z^2}\equiv |2,0\rangle $ and $|B\rangle\equiv(d_{x^2-y^2}- i \tau_z d_{xy})/\sqrt{2}\equiv |2,-2\tau_z\rangle$ orbitals centered on a single Mo atom.  

\begin{figure}
\begin{center}
\includegraphics[width=0.49\textwidth]{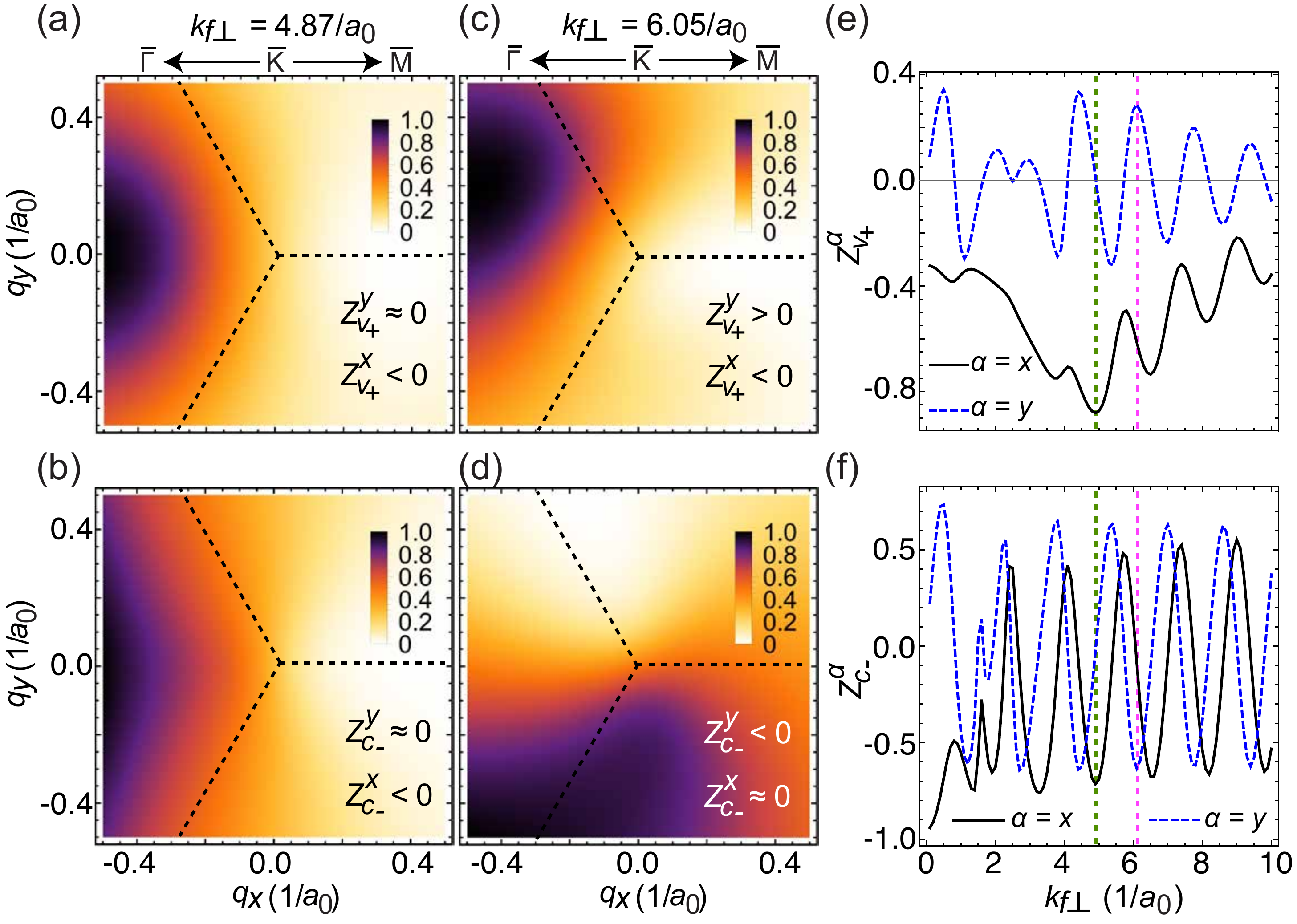}
\caption{(a)-(d) Color map of $|{\cal M}_n|^2$ in the BL MoS$_2$ (a) VB and (b) CB for (a)-(b) $k_{f\perp} \approx 4.87/a_0$ and the (c) VB and (d) CB for (c)-(d) $k_{f\perp} \approx 6.05/a_0$. In each panel the sign of the masking parameter $Z^{\alpha}_n$ evaluated along the direction $q_\alpha$ and in the CB ($n = c_-$) and VB ($n = v_+$) is provided. The magnitude of $|{\cal M}_n|^2$ has been normalized to the maximum value in the plotted range. (e)-(f) Masking parameter calculated along $q_x$ ($\alpha = x$) and $q_y$ ($\alpha = y$) as a function of $k_{f\perp}$ for the (e) VB and (f) CB. We set $\tau_z=1$, $q_c=0.5/a_0$ and average over the spin, $s_z$. The vertical dashed lines indicate the $k_{f\perp}$ values used for the plots in (a)-(d).}
\label{fig:probe_bl}
\end{center}
\end{figure}

In the BL TMDC model, the square modulo of the matrix element reads 
\begin{align}\label{Eq:BLtmMn}
|{\cal M}_{n}|^2 & = |{\cal M}^+_n|^2+|{\cal M}^-_n|^2 
\nonumber\\&+ 2{\rm Re}[{\cal M}^{+\ast}_n {\cal M}^{-}_n]\cos(ck_{f\perp})
\nonumber\\&+2{\rm Im}[{\cal M}^{+\ast}_n {\cal M}^{-}_n]\sin(ck_{f\perp}).
\end{align}
Here, we find two sets of interference effects: One originates from the Mo $d$-orbitals within each layer and can accordingly be referred to as an {\it intra-layer interference} effect. It is given by the first line in Eq. (\ref{Eq:BLtmMn}). The other comes from the layer degree of freedom as seen by the last two lines in Eq. (\ref{Eq:BLtmMn}) and is therefore referred to as {\it inter-layer interference}.

The result of a numerical evaluation of Eq. (\ref{Eq:BLtmMn}) is presented in Figs.~\ref{fig:probe_bl}(a)-(b) for $k_{f\perp}\approx 4.87/a_0$ and Figs.~\ref{fig:probe_bl}(c)-(d) for $k_{f\perp}\approx 6.05/a_0$ in the top-layer VB ($n=v_+$) and bottom-layer CB ($n=c_-$), which constitute the VBM and CBM, respectively. Indeed, the masking effect appears to be considerably different from that of SL MoS$_2$ and varies strongly along both the $q_x$- and $q_y$-direction and this variation depends also on $k_{f\perp}$. A calculation of the masking parameter for different $k_{f\perp}$ reveals an oscillatory behavior as shown in Figs.~\ref{fig:probe_bl}(e) and \ref{fig:probe_bl}(f) for the VB and CB, respectively, which reflects the inter-layer $\cos(ck_{f\perp})$ and $\sin(ck_{f\perp})$ terms in Eq. (\ref{Eq:BLtmMn}). Such behavior was not observed in a previous study of the matrix elements in BL graphene in Ref. \citenum{mucha_prb_2008} because it was assumed that $k_{f\perp} c\ll1$. Our result presented here indicate a significant photon energy dependence of the photoemission matrix elements for ARPES from BL TMDCs. Since a bulk TMDC consists of the same unit cell as the BL investigated here, the photon energy dependence of the matrix element can be expected to behave in a very similar way in the bulk.

\subsection{Calculation of ARPES intensity}
\begin{figure}
\begin{center}
\includegraphics[width=0.49\textwidth]{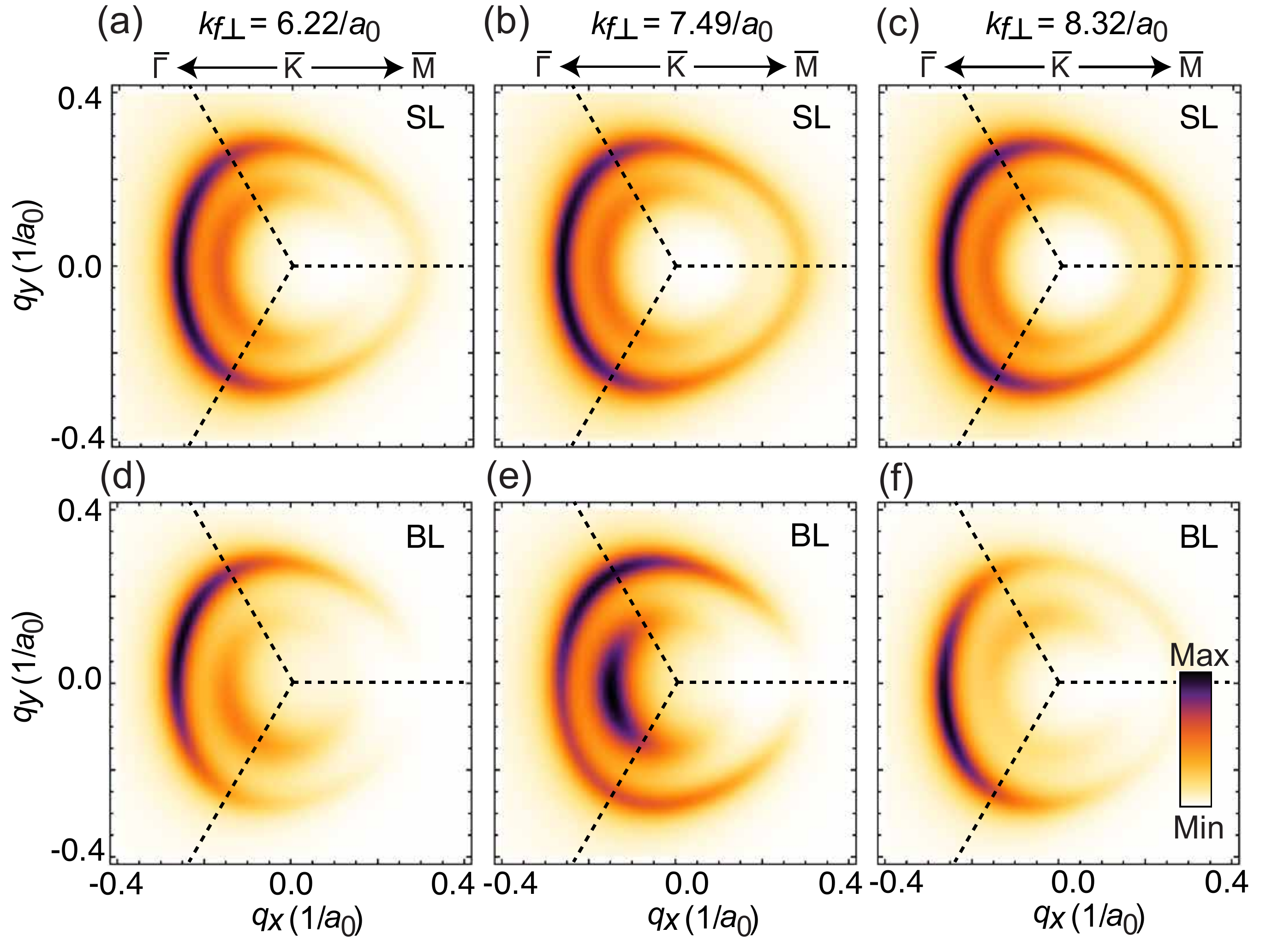}
\caption{Calculated photoemission intensity cuts in the VB of (a)-(c) SL MoS$_2$ and (d)-(f) BL MoS$_2$ at a constant energy of $E-E_{\rm VBM}=-0.24$~eV at the values of $k_{f\perp}$ stated above each column.}
\label{fig:intensity}
\end{center}
\end{figure}
The photoemission intensity of SL and BL MoS$_2$ is numerically calculated for the VB using the expression
\begin{align}
{\cal I}_n
\propto  |{\cal M}_n(k_{f\perp};{\bm q},\tau_z,s_z)|^2 {\cal A}_n(E,{\bm q},\tau_z,s_z),
\end{align}
where $f_{FD} = 1$, since we are considering the occupied states. The spectral function is expressed as ${\cal A}_n(E,{\bm q},\tau_z,s_z) = \pi^{-1}\Gamma/(\left[E-E_n({\bm q},\tau_z,s_z)\right]^2 + \Gamma^2)$, where we set $\Gamma = 0.05$~eV and the bare dispersion $E_n({\bm q},\tau_z,s_z)$ is obtained by diagonalizing the corresponding $\bm k\cdot\bm p$  Hamiltonian.

The $\bm q$-dependent intensity in the VB at an energy of $E-E_{\rm VBM}=-0.24$~eV is presented in Fig.~\ref{fig:intensity} for SL and BL MoS$_2$. We use $k_{f\perp}$ values of $6.22/a_0$, $7.49/a_0$ and $8.32/a_0$ because these correspond to the photon energies of 49~eV, 65~eV and 80~eV used for the ARPES measurements in Fig. \ref{fig:expFig}, if an inner potential given by the bulk MoS$_2$ value, $V_0 = 12$~eV, is assumed \cite{Boker:2001}. A comparison between these values has a rather high level of uncertainty because of the ambiguity of $V_0$ and the choice of tight-binding parameters in the modeling. However, we observe excellent agreement between the measured ARPES intensity of SL MoS$_2$ in Fig. \ref{fig:expFig}(a) and the calculations in Figs.~\ref{fig:intensity}(a)-(c) where a strong masking effect is observed along $q_x$, reducing the overall intensity towards the outer BZ. In BL MoS$_2$ we sum over the spin index and find a qualitative agreement between the calculations in Figs.~\ref{fig:intensity}(d)-(f) and the data in Figs. \ref{fig:expFig}(b)-(c), as the intensity in the two trigonally warped contours is seen to concentrate along the BZ edges for some $k_{f\perp}$-values. This behavior emerges from the inter-layer interference effect that leads to a $k_{f\perp}$-dependent masking effect along $q_y$.

\section{Light-induced conduction band population}
In this section we investigate the ${\bm q}$-dependent CB intensity that can be measured in TR-ARPES performed on TMDCs as shown in Ref. \citenum{ourPRL}. An initial optical excitation by an intense laser pulse with a photon energy close to the TMDC direct gap at \kbar~leads to direct interband transitions between VB and CB states. The resulting excited state is then probed by photoemission, such that the TR-ARPES intensity from the CB will depend on both the photoemission matrix element, which was explored in the previous section, and the momentum dependence of the transient population. The latter can be determined by solving the semiconductor Bloch equations in the dipole approximation for the light-matter interaction, which we consider in the following. The time- and polarization-dependent photoemission intensity in the excited state is then written 
\begin{align}\label{eq:tr_arpes}
{\cal I}_n(E,{\bm q},\theta,t) &\propto  |{\cal M}_n({\bm q})|^2 
{\cal A}_n(E,{\bm q}) f_n({\bm q},\theta,t)
\end{align}
where $f_n({\bm q},\theta,t)$ is the transient CB population and $\theta$ determines the direction of the electric field polarization vector, $\hat{\bm \epsilon}(\theta)=\cos\theta\hat{\bm x} + i \sin\theta \hat{\bm y}$, for the optical excitation. Note that time-dependent many-body effects such as screening induced band renormalization and electron-phonon interactions lead to a time-dependent spectral function \cite{Andreatta:2019,Ulstrup:2016,johannsen:2013}, which we neglect in the analysis presented here. Our results are thus restricted to the situation in the initially excited state before any dephasing and relaxation processes occur.

\subsection{The semiconductor Bloch equations}

We consider the following Hamiltonian for the light-matter interaction in the band basis:
\begin{align}
\hat {\cal H} = \sum_n E^c_n\hat c^\dagger_n  \hat c_n 
+\sum_{m} E^v_m  \hat v^\dagger_m    \hat v_m + \sum_{nm} \hbar\omega_{R;nm} \hat c^\dagger_n \hat v_m + c.c.
\end{align}
Note that $\hat{c}^\dagger_n$ ($\hat{v}^\dagger_m$) stands for the creation operator for the $n$th conduction ($m$th valence) band. The parameter $\omega_{R;nm}= \bm {\mathcal E}\cdot {\bm d}_{nm}/\hbar $ is the Rabi frequency that quantifies the light-matter interaction in which ${\bm d}_{nm} = \langle n|(-e\bm r)|m \rangle$ is the dipole moment matrix element. The laser pulse is modeled as a classical electric field $\bm {\mathcal E}(\theta,t) = \hat {\bm \epsilon}(\theta) \cos(\omega_0 t) \exp\{-{t^2}/{\tau^2_0}\}$ where $\tau_0$ is the pulse duration and $\omega_0$ is the pump laser frequency.  
The CB (VB) population is written as $f^c_n= \langle  \hat c^\dagger_n \hat c_n\rangle$ ($f^v_m= \langle  \hat v^\dagger_m \hat v_m\rangle$) and is determined by solving the semiconductor Bloch equations within the rotating wave approximation (RWA), which implies $e^{\pm i\omega_0 t}\cos(\omega_0 t) \to 1/2$ for a high frequency \cite{Haug_Koch,Rossi}: 
\begin{align}
&\frac{\partial f^c_n }{\partial t} = - 2 \sum_{m} {\rm Im}[ \omega_{R;nm}  p^\ast_{nm} ], \label{eq:fc} 
\\
&\frac{\partial f^v_m }{\partial t} =  +2 \sum_{n} {\rm Im}[ \omega_{R;nm}  p^\ast_{nm} ], \label{eq:fv} 
\\
& \left[\frac{\partial}{\partial t} +i\nu_{nm}+ \frac{1}{\tau_p}\right] p_{nm} = -i \omega_{R;nm} (f^c_n -f^v_m). \label{eq:p} 
\end{align}
Note that, as before, $n=\pm$ and $m=\pm$ correspond to the sub-band label in the CB and VB, respectively. 
The parameter $\tau_p$ stands for the relaxation of interband polarization, $p_{nm} = e^{-i\omega_0 t}\langle  \hat c^\dagger_n \hat v_m\rangle $. 
The detuning parameter reads $\nu_{nm}(\bm k) = [E^c_{n}(\bm k)-E^v_{m}(\bm k)]/\hbar-\omega_0$ and the Rabi frequency is given by 
\begin{align}
\omega_{R;nm} (\bm k,\theta,t) =\frac{{\cal E}_0 \hat{\bm \epsilon}(\theta)\cdot {\bm d}_{nm}(\bm k)}{2\hbar} e^{-\frac{t^2}{\tau^2_0}},
\end{align}
where the factor of $1/2$ originates from the RWA. Note that the dipole moment matrix element between Bloch states, $|\psi^{c/v}_j(\bm k)\rangle = e^{i\bm k\cdot\bm r}|u^{c/v}_{j}(\bm k)\rangle$, can be evaluated as
\begin{align}
{\bm d}_{nm}(\bm k)  
=\frac{i e\hbar}{E^c_{n}(\bm k)-E^v_{m}(\bm k)} 
\langle u^c_{n}(\bm k) |{\bm \nabla}_{\bm k}\hat{\cal H}(\bm k)| u^v_{m}(\bm k)\rangle,
\end{align}
where we applied the identity $[\hat{\bm r},\hat{\cal H}(\bm k)] = i {\bm \nabla}_{\bm k} \hat{\cal H}(\bm k)$. 
Therefore, the Rabi frequency reads as 
\begin{align}
\omega_{R;nm} (\bm k,\theta,t) =i 
\frac{e {\cal E}_0 e^{-\frac{t^2}{\tau^2_0}}}{2(E^c_{n}(\bm k)-E^v_{m}(\bm k))} {\cal M}_{nm}(\bm k,\theta),
\end{align}
where the central quantity is the velocity matrix element ${\cal M}_{nm}(\bm k,\theta) = \langle u^c_{n}(\bm k) |\hat{\bm \epsilon}(\theta)\cdot{\bm \nabla}_{\bm k}\hat{\cal H}| u^v_{m}(\bm k)\rangle$, which depends on the polarization vector of the pump pulse. 
\par
In a perturbative treatment of the strength of the electric field, we have the following linearized equation of motion for $p_{nm}$: 
\begin{align}
\left[\frac{\partial}{\partial t} +i\nu_{nm} + \frac{1}{\tau_p}\right] p_{nm}(t)= i \omega_{R;nm} (t).
\end{align}
Note that we have considered $f^c_n=1-f^v_m=0$ in the absence of the external field.  
Using $p_{nm}(0)=0$, we find
\begin{align}
p_{nm}(t)= \int^t_0 dt' \omega_{R;nm} (t') G_{nm}(t'-t),
\end{align}
where $G_{nm}(t)=i e^{-i(\nu_{nm}-i/\tau_p)t}$. We plug the above relation into Eq.~(\ref{eq:fc}) and by 
considering $f^c_n(0)=0$, we find the total excited state population in the CB
\begin{align}
f^{exc}_n
= -2 \sum_m  {\rm Im} \left [ \int^\infty_{0} dt \omega_{R;nm} (t) p^\ast_{nm}(t)\right].
\end{align}
This expression can be simplified as follows 
\begin{align}
f^{exc}_n  
= \frac{(e{\cal E}_0)^2}{2} \sum_m  \frac{|{\cal M}_{nm}|^2}{(E^c_n-E^v_m)^2}    K_{nm}(\nu_{nm},\tau_0,\tau_p)
\end{align}
where 
\begin{align}
K_{nm}(\nu_{nm},\tau_0,\tau_p) &= \int^{\infty}_{0}  dt \int^{t}_{0} dt'  e^{-\frac{t^2+t'^2}{\tau^2_0}} 
{\rm Im}[G_{nm}(t'-t)].
\end{align}
The integral over $t'$ can be solved analytically, leading to
\begin{align}
K_{nm}(\nu_{nm},\tau_0,\tau_p) &= \int^{\infty}_{0}  dt~g(t,\nu_{nm},\tau_0,\tau_p)
\end{align}
where 
\begin{align}
&g(t,\nu_{nm},\tau_0,\tau_p)=\frac{\sqrt{\pi } \tau _0}{4} e^{-\frac{\tau _0^2\tilde\nu^{\ast2}_{nm}}{4}} e^{-it \tilde\nu^\ast_{nm}-\frac{t^2}{\tau _0^2}}
\nonumber\\&
\times\left[ S^\ast(t, \tilde\nu_{nm},\tau_0)
   +e^{i \nu_{nm}  \tau_0 \left(\frac{\tau _0}{\tau _p}+2 \frac{t}{\tau_0}\right)} S(t, \tilde\nu_{nm},\tau_0)
\right].
\end{align}
Note that $\tilde\nu_{nm}=\nu_{nm}-i/\tau _p$ and $S(t, \tilde\nu_{nm},\tau_0)=\text{erf} (\frac{t}{\tau _0}+i\frac{\tau _0 \tilde\nu_{nm}   }{2}    )- 
   \text{erf} (\frac{i}{2} \tau _0 \tilde\nu_{nm} )$ in which ${\rm erf}(x)$ is the error function. For the resonance condition, i.e. $|\tilde\nu_{nm}|\ll 1/\tau_0$, we can approximate $g\approx (\sqrt{\pi}\tau_0/2) e^{-t^2/\tau^2_0} {\rm erf}(t/\tau_0)$,  which implies 
$K_{nm} \approx \pi \tau^2_0/8$. Moreover, for a large bandgap system like MoS$_2$, we have  $E^c_n-E^v_m\approx E_g$ which implies the following relation around the valley points
\begin{align}
f^{exc}_n
\approx \frac{\pi \tau^2_0(e{\cal E}_0)^2}{16 E^2_g} \sum_m  |{\cal M}_{nm}|^2.
\end{align}
Therefore, to calculate the transient population we need to evaluate the velocity matrix elements as a function of ${\bm q}$, which is discussed in the following subsection.
\subsection{Evaluation of velocity matrix elements}
Since the optical interband transitions are spin ($s_z$) and valley ($\tau_z$) conserving we can define the velocity matrix element associated with the pump pulse as follows
\begin{align} 
{\cal M}_{nm}(\bm q,\tau_z,s_z) =    \langle u^{c}_n | \hat {\bm \epsilon}(\theta)\cdot{\bm \nabla}_{\bm q} \hat {\cal H}(\bm q,\tau_z,s_z) |u^{v}_m \rangle,
\end{align}
where for shorthand notation we drop the $({\bm q},\tau_z,s_z)$-dependence of the wave functions. We consider an elliptically polarized pump pulse, where $\theta =0$ and $\theta =\pi/2$ correspond to linear horizontal and vertical polarizations, while the cases of $\theta=\pm\pi/4$ correspond to left- ($-$) and right-handed ($+$) circular polarizations.  
For a given optical transition $m\to n$, we formally write
${\cal M}_{nm} = M^{nm}_x \cos\theta+i M^{nm}_y \sin\theta$ in which the velocity matrix element component reads 
$M^{nm}_\alpha=  \langle u^c_n   | \partial_{q_\alpha} \hat {\cal H} | u^v_m  \rangle$. The absolute value square then follows 
\begin{align}
|{\cal M}_{nm}|^2 =  |M^{nm}_0|^2 \left(1+f^{nm}_{\rm circ} \sin(2\theta) 
+ f^{nm}_{\rm lin} \cos(2\theta)  \right)
\end{align}
where $2|M^{nm}_0|^2 =|M^{nm}_x|^2+|M^{nm}_y|^2$ and
\begin{align}
&f^{nm}_{\rm circ}=\frac{2 {\rm Im}[M^{nm}_x M^{nm\ast}_y]}{|M^{nm}_x|^2+|M^{nm}_y|^2}~,
\\
&f^{nm}_{\rm lin}=\frac{|M^{nm}_x|^2-|M^{nm}_y|^2}{|M^{nm}_x|^2+|M^{nm}_y|^2}~.
\end{align}
The term proportional to $\sin(2\theta)$ leads to a {\it circular dichroism effect}, which is normally exploited to generate a valley-polarization in the SL TMDCs \cite{zengvalley2012,makcontrol2012,Cao_2012} because $\sin(2\theta) = \pm1$ for $\theta = \pm \pi/4$ where $+/-$ stands for right/left-handed circular polarization. On the other hand, the $\cos(2\theta)$ term corresponds to a {\it linear dichroism effect} since $\cos(2\theta)$ is equal to 1 (-1) when $\theta$ is 0 ($\pi/2$).  
\par
It is interesting to note that, because of the proportionality $f_{\rm circ}(\bm q)\propto {\rm Im}[M_x M^\ast_y]$, this term is closely related to the Berry curvature $\Omega(\bm q)$, which is given by  \cite{chang_prb_1996,yao_prb_2008}
\begin{align}
\Omega_i({\bm q}) = -  \sum_{j\neq i}\frac{2{\rm Im} [M^{ij}_x({\bm q}) M^{ij\ast}_y ({\bm q})]}{(E_i({\bm q})-E_j ({\bm q}))^2},
\end{align}
such that $\Omega(\bm q)$ can be obtained by extracting $f_{\rm circ}(\bm q)$ in a TR-ARPES measurement \cite{ourPRL}. The spin-averaged Berry curvature is zero if the system is invariant under both spatial inversion and time-reversal symmetries. This implies that circular dichroism is absent in a BL TMDC owing to the centrosymmetric structure of the system. 

\begin{figure}
\begin{center}
\includegraphics[width=0.49\textwidth]{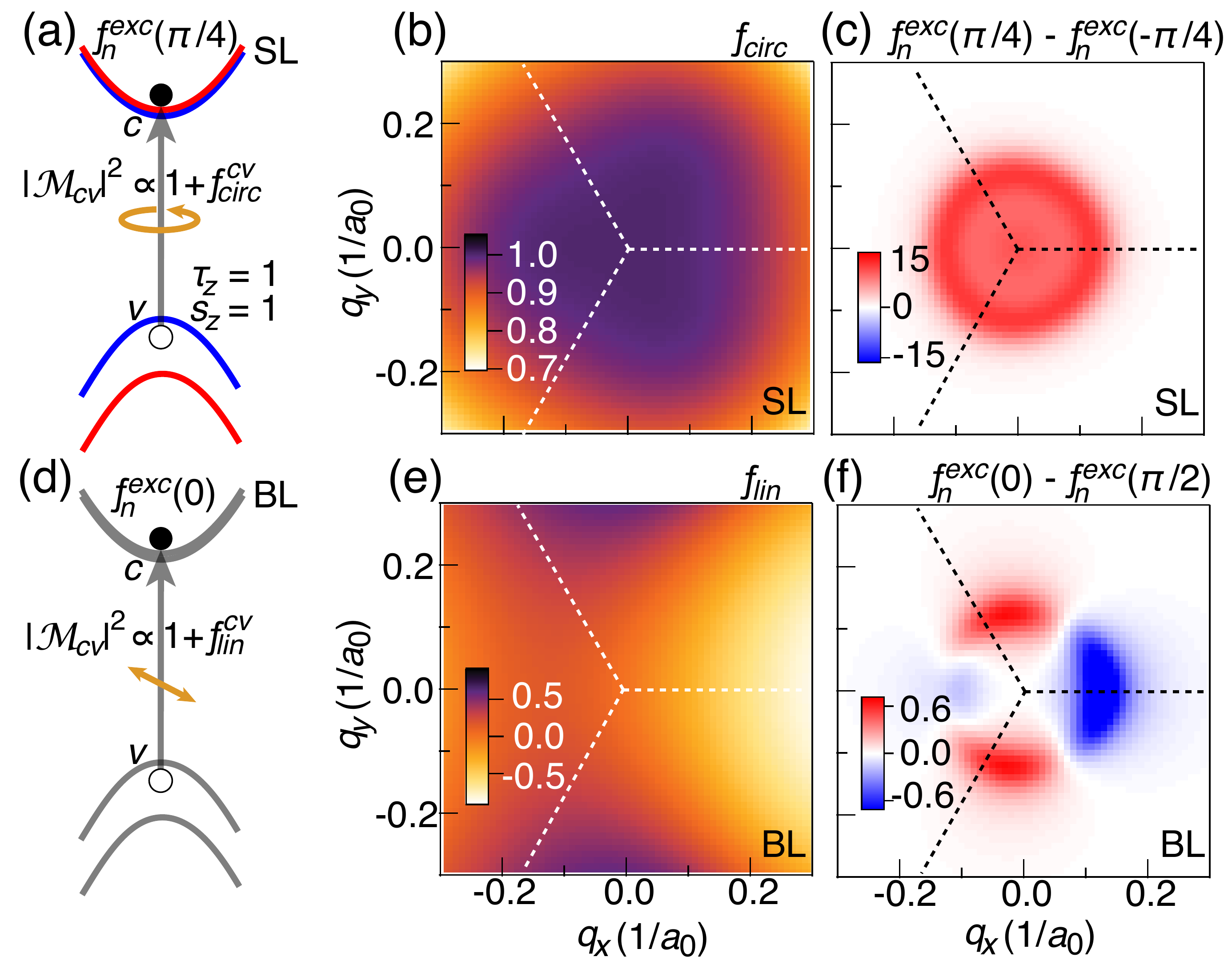}
\caption{(a) Sketch of electron (filled circle) and hole (open circle) excitation with a circularly polarized ($\theta = \pi/4$) light pulse at the \kbar-valley ($\tau_z = 1$) in a SL TMDC. The arrow indicates a transition from the VB ($v$) to the CB ($c$) for a pulse energy resonant with the direct band gap. The velocity matrix element $|{\cal M}_{cv}|^2$ governing the transition is indicated for $\theta = \pi/4$. Blue (red) curves indicate spin up (down), corresponding to $s_z = 1$. (b) ${\bm q}$-dependent circular dichroism component $f_{\rm circ}$ of SL MoS$_2$ corresponding to the spin up VB and CB states. (c) Difference of excited state population $ f^{exc}_n$ in SL MoS$_2$ between optical excitations with circular left- and right-polarizations. (d) Similar sketch as in (a) for a BL TMDC excited with a linearly polarized ($\theta = 0$) pulse along with the corresponding expression for $|{\cal M}_{cv}|^2$. (e) ${\bm q}$-dependent linear dichroism component $f_{\rm lin}$ of BL MoS$_2$ averaged over spin due to the spin degeneracy in BL MoS$_2$. (f) Difference of excited state population $ f^{exc}_n$ in BL MoS$_2$ between optical excitation with linear vertical and horizontal polarizations. All calculations were performed by using $\tau_0=30$~fs, $\tau_p\approx22$~fs, ${\cal E}_0=0.87$~V/nm and $\hbar\omega_0=2$~eV.}
\label{fig:pump}
\end{center}
\end{figure}

By diagonalizing the two-band Hamiltonian given in Eq. (\ref{eq:Ham_SL}), we calculate $M_\alpha$ and thereby obtain the velocity matrix element, ${\cal M}_{cv}$, for a transition from the VB to the CB for a given set of spin and valley indices in SL MoS$_2$, as sketched in Fig. \ref{fig:pump}(a) for $\theta = \pi/4$. The numerical result for the $f_{\rm circ}$ component is shown in  Fig. \ref{fig:pump}(b) for $s_z = 1$ and $\tau_z = 1$. This term is strong and nearly isotropic around \kbar. By solving the Bloch equations given in Eqs.~(\ref{eq:fc})-(\ref{eq:p}) we obtain the transient population in the CB and find a nearly uniform circular dichroism effect in momentum space when calculating the difference $f^{exc}_n(\theta=\pi/4) - f^{exc}_n(\theta=-\pi/4)$, which is shown in Fig. \ref{fig:pump}(c).

It is instructive to give analytical expressions for both $f_{\rm circ}$ and $f_{\rm lin}$ for small $q$ in a SL TMDC.  Neglecting the spin-orbit coupling, we have $|M_0|^2 \approx v^2$ with $v=a_0 t_1/\hbar$ and
\begin{align}
&f_{\rm circ} \approx  \tau_z  \left[1 - 2 R^2 (a_0q)^4 \right ] + 8R \frac{t_2}{t_1} (a_0 q)^3 \cos(3\phi)~,
\\
&f_{\rm lin} \approx R (a_0 q)^2\cos(2\phi)  + 4\tau_z\frac{t_2}{t_1} (a_0 q) \cos(\phi)
\label{eq:flin}~, 
\end{align}
where $R=(t^2_1-2 E_0 E_g)/E^2_g$ and $\phi = \arctan(q_y,q_x)$. The linear dichroism term is non-zero only at finite $q$ where the symmetry between the $q_x$- and the $q_y$-direction is broken. 

\par
By diagonalizing the four-band Hamiltonian of BL MoS$_2$ given in Eq.~(\ref{eq:BL_4b}), we calculate the velocity matrix elements and extract the spin-averaged linear dichroism following excitation with a linearly polarized pump pulse as sketched for $\theta = 0$ in Fig. \ref{fig:pump}(d). The $f_{\rm lin}$ component shown in Fig.~\ref{fig:pump}(e) is highly anisotropic, indicating a strongly $q$-dependent transient population when excited with linearly polarized light, even in an inversion symmetric BL TMDC. This is more clearly seen in Fig. \ref{fig:pump}(f) via the difference in transient population induced using excitations with linear horizontal and vertical polarizations, $f^{exc}_n(\theta=0) - f^{exc}_n(\theta=\pi/2)$, which is again found by solving the Bloch equations given in Eqs.~(\ref{eq:fc})-(\ref{eq:p}). Since circular dichroism is absent in a BL TMDC the optical response in a TR-ARPES experiment probing a single $(E,k)$-cut for different pump pulse polarizations will be dominated by the noticeably strong linear dichroism around \kbar~in Fig. \ref{fig:pump}(f) \cite{ourPRL}. 

\section{Summary}
We have calculated the photoemission matrix elements for SL MoS$_2$ around the \kbar~valley using a single free-electron final state approximation and a two-band $\bm k\cdot\bm p$ Hamiltonian that includes trigonal warping effects. The model was extended to BL MoS$_2$ using a four-band Hamiltonian. In photoemission from the VB states in SL MoS$_2$ we find that intra-layer interference arising from the transition metal $d$ orbitals causes a suppression of photoemission intensity towards the higher Brillouin zones. An additional inter-layer interference in BL MoS$_2$ leads to a complex redistribution of intensity that strongly depends on the photon energy in an ARPES experiment. We note that while MoS$_2$ was used as an example, the modeling applies to other semiconducting TMDCs, merely requiring the adaptation of their $\bm k\cdot\bm p$ parameters. 

In order to describe the intensity of the CB states in a TR-ARPES experiment, we have numerically solved the semiconductor Bloch equations and evaluated the interband velocity matrix elements, describing transitions from the occupied VB to the unoccupied CB based on an optical excitation with tunable polarization. In SL TMDCs the transient population in the CB exhibits a near uniform circular dichroism in momentum space while in BL TMDCs this effect is absent, thereby leaving behind a highly momentum-dependent linear dichroism effect at finite ${\bm q}$.

The differences in photoemission intensity and transient state behavior that we have found here in SL and BL TMDCs underline the crucial role of both orbital and layer degrees of freedom when interpreting (TR) ARPES spectra from these materials. Our results for the photoemission and inter-band matrix elements will facilitate the analysis of dichroism and Berry curvature in the TMDCs, as well as help deconvolve matrix element effects from many-body interactions in the spectral function of the materials.
 
\section{Acknowledgements}

We gratefully acknowledge funding from VILLUM FONDEN through the Young Investigator Program (Grant. No. 15375) and the Centre of Excellence for Dirac Materials (Grant. No. 11744), the Danish Council for Independent Research, Natural Sciences under the Sapere Aude program (Grant No. DFF-4002-00029 and DFF-6108-00409) and the Aarhus University Research Foundation. H.R. acknowledges the support from the Swedish Research Council (VR 2018-04252).

\end{document}